\documentclass[pre]{revtex4}

\usepackage{amsmath}
\usepackage{amsfonts}
\usepackage{hyperref}
\usepackage{graphicx}

\usepackage[T1]{fontenc}

\renewcommand{\d}{\mathrm{d}}%

\begin{document}

\title{On the Bessel solution of Kepler's equation}

\author{Riccardo Borghi}
\affiliation{Dipartimento di Ingegneria Civile, Informatica e delle Tecnologie Aeronautiche, Universit\`{a} ``Roma Tre'', Via Vito Volterra 62, I-00146 Rome, Italy}

\begin{abstract}
Since its introduction in 1650, Kepler's equation has never ceased to fascinate mathematicians, scientists, and engineers.
Over the course of five centuries, a large number of different solution strategies have been devised and implemented. Among them, the one originally proposed by J. L. Lagrange and later by F. W. Bessel still 
continue to be a source of mathematical treasures. 
Here, {{the Bessel solution of the elliptic Kepler equation}} 
is explored from a new perspective offered by the theory of the {Stieltjes series}. 
In particular, it has been proven that a complex Kapteyn series obtained directly by the Bessel expansion is a Stieltjes series. This mathematical result, to the best of our knowledge, is a new integral 
representation of the KE solution.  
Some considerations on possible extensions of our results to more general classes of the Kapteyn series are also presented.
\end{abstract}

\maketitle

\section{Introduction}
\label{Sec:Intro}

There is no other equation in the history of science that boasts more solution strategies than the famous {Kepler equation} (hereafter KE). 
Since 1650, a huge number of different strategies and algorithms have been devised and implemented to deal with KE. 
The classic textbook by Colwell~\cite{Colwell/1993} lists and briefly describes most of these methods. However, since 1993 (the year of publication of Colwell's book), this list has  
has grown so rapidly that it is difficult to account for all of the strategies that have been developed.
Just to gain an idea, quoting some of the papers that have been published on the subject in the last {five} 
 years would be sufficient~\cite{
Ibrahim/Saleh/2019,%
Calvo/Elipe/Montijano/Randez/2019,%
Tommasini/Olivieri//2020,%
Abubekerov/Gostev/2020,%
Sacchetti/2020,%
Tommasini/Olivier/2020,%
Zechmeister/2021,%
Gonzalez/Hernandez/2021,%
Tommasini/2021,%
Philcox/Goodman/Slepian/2021, %
Tommasini/Olivieri/2022, 
Zhang/Bian/Li/2022,%
Zhou/Lim/Zhong/2023,%
Vavrukh/Dzikovskyi/Stelmakh/2023,
Calvo/Elipe/Randez/2023,%
Brown/2023}. 

In this paper, we focus on a semi-analytic approach to the solution of KE,  
originally conceived by J. L. Lagrange and later finalised by F. W. Bessel via a Fourier series expansion,
in which his famous functions appeared. The~whole third chapter of~\cite{Colwell/1993} is devoted to an account 
of the history of the infinite series solutions of KE, starting with an important theorem proven by Lagrange in 1770,  
to the series expansion proposed by T. Levi-Civita in 1904.  
Today, the~Bessel solution of KE has been abandoned for practical, computational purposes, although~it continues to arouse some interest, especially regarding its mathematical properties. The~Bessel solution is the first example of a class of expansions called the { {Kapteyn series}}  {} 
(hereafter KS)~\cite{Kapteyn/1893}, which has gained considerable importance in mathematics and theoretical physics, even in recent times~\cite{%
Dominici/2007,Lerche/Tautz/2007,Pogany/2007,Lerche/Tautz/2008,Lerche/Tautz/Citrin/2009,Eisinberg/2010,Tautz/Lerche/Dominici/2011,Baricz/Jankov/Pogany/2011,Nikishov/2014,Baricz/Dragana/Masirevic/Pogany/2017,Xue/Li/Man/Xing/Liu/Li/Wu/2019,Bornemann/2023}. 

In the present paper, the~Bessel solution of KE is studied from the perspective of the theory of so-called { {Stieltjes functions}}~\cite{Bender/Orszag/1978,Baker/Graves-Morris/1996,Widder/1938}. 
In particular, two main results are established: (i) a Kapteyn series obtained by a simple ``complexification'' of the Bessel series is proven to be Stieltjes
and (ii) thanks to such a (constructive) proof, to~our knowledge, a new integral representation of the KE solution will be obtained.
We also hope that, under~this perspective, the~availability of a well-established convergence theory for Stieltjes series based on the Pad\'e approximant~\cite{Baker/Graves-Morris/1996,Brezinski/1996}, could provide new life to the Bessel solution, even as a computational~tool.


\section{Bessel' Solution of Elliptic Kepler's~Equation}
\label{Sec:TheoAnal}

\begin{figure}[!ht]
\includegraphics[width=6cm]{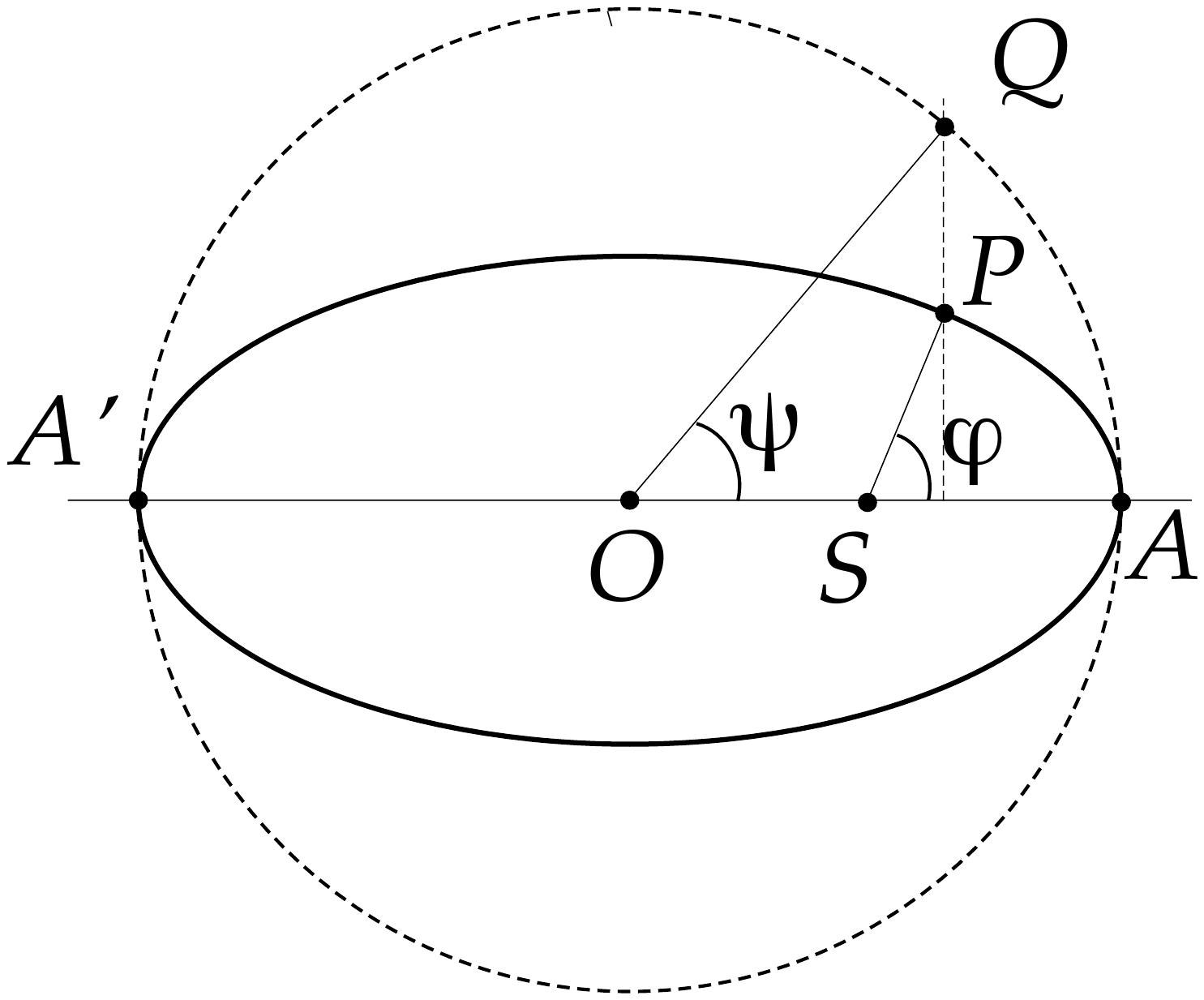}
\caption{The geometry of Kepler's~Equation.}
\label{Fig:KEGeometry}
\end{figure}

Consider the motion of a planet around the star $S$ along an elliptic orbit with eccentricity $\epsilon$, as~sketched in Figure~\ref{Fig:KEGeometry}~\cite{Orlando/Farina/Zarro/Terra/2018}.

Suppose $P$ is the position of the planet at time $t$, which is measured by assuming 
the pericenter $A$  as the initial (i.e., at~$t=0$) position (motion is counter-clockwise). Upon~introducing the mean  angular speed $\omega=2\pi/T$, with~$T$ 
being the orbital period, the~so-called { {mean anomaly}}, say $M$, is defined by $M=\omega t$.
To retrieve the position of the planet along the orbit at time $t>0$, the~angle $\psi$ is first evaluated by solving the following trascendental  {equation:} 
%
\begin{equation}
\label{Eq:KE.1}
M\,=\,\psi\,-\,\epsilon\,\sin\,\psi\,,
\end{equation}
which provides the position of point $Q$ along the circumscribed circle (dotted in the figure). The planet's position $P$ 
is then immediately obtained through the geometrical construction sketched in Figure~\ref{Fig:KEGeometry}.
Equation~(\ref{Eq:KE.1}) is called the {elliptic} KE.

The Bessel solution of Equation~(\ref{Eq:KE.1}) is provided by~\cite{Colwell/1993} (Ch.~3)
\begin{equation}
\label{Eq:KE.2}
\begin{array}{l}
\displaystyle
\psi\,=\,M\,+\,S(\epsilon;M)\,,
\end{array}
\end{equation}
where function $S(\epsilon;M)$ represents the Fourier series 
\begin{equation}
\label{Eq:KE.3}
\begin{array}{l}
\displaystyle
S(\epsilon;M)\,=\,\sum_{n=1}^\infty\,\frac{2\,J_n(n\,\epsilon)}n\,\sin\,nM\,.
\end{array}
\end{equation}
 {Although} 
 the  series converges  for any  $\epsilon\in[0,1)$, its convergence turns out to be extremely slow, especially when $\epsilon \to 1$. 
This fact, together with the considerable number of Bessel function evaluations to be implemented, unavoidably resulted in Equations~(\ref{Eq:KE.2}) and~(\ref{Eq:KE.3})  being
abandoned as far as the practical use of KE was~concerned. 

Nevertheless,  the~Bessel series expansion~(\ref{Eq:KE.3}) remained a subject of considerable 
interest in both mathematics and theoretical physics. The~scope of the present paper is to explore why this occurred.
To this end, we introduce the complex function $\mathbb{S}(\epsilon;M)$, defined through the  Kapteyn series
\begin{equation}
\label{Eq:KE.4}
\begin{array}{l}
\displaystyle
\mathbb{S}(\epsilon;M)\,=\,\sum_{n=1}^\infty\,\frac{2\,J_n(n\,\epsilon)}n\,\exp(\mathrm{i}nM)\,,
\end{array}
\end{equation}
in such a way that $S=\mathrm{Im}\{\mathbb{S}\}$.
The  convergence of the series~(\ref{Eq:KE.4}) can be proven, for~example, by~invoking the following  Bessel function asymptotics, valid for  
$\epsilon < 1$~\cite{Watson/1944} (Sec.~8.4):
\begin{equation}
\label{Eq:KE.4.1}
\begin{array}{l}
\displaystyle
J_n(n\,\epsilon)\,\sim\,\frac 1{\sqrt{2\pi\chi}}\,\frac{\exp(n\lambda)}{\sqrt n}\,,\qquad\quad  n\to\infty\,,
\end{array}
\end{equation}
where $\chi=\sqrt{1-\epsilon^2}$ denotes the ellipse's aspect ratio and  $\lambda$ is a { {negative}} parameter, provided  by
\begin{equation}
\label{Eq:KE.4.2}
\begin{array}{l}
\displaystyle
\lambda=\chi\,+\,\dfrac 12\,\log\dfrac{1-\chi}{1+\chi}\,.
\end{array}
\end{equation}
%
 {The} parameter $\lambda$ will play a key role in the rest of the~paper. 

After substituting Equation~(\ref{Eq:KE.4.1}) into Equation~(\ref{Eq:KE.4}) and upon introducing the complex parameter 
$z=\exp(\lambda+\mathrm{i}M)$, it can immediately be seen that the single terms of the series~(\ref{Eq:KE.4}) asymptotically approach 
those of the paradigmatic model series
\begin{equation}
\label{Eq:KE.6}
\begin{array}{l}
\displaystyle
\sum_{n=1}^\infty\,\frac{z^n}{n^{3/2}}\,,
\end{array}
\end{equation}
which, for~$|z|<1$, is nothing but the Dirichlet series representation of the so-called polylogarithm function $\mathcal{L}_{3/2}(z)$, where 
\begin{equation}
\label{Eq:KE.6.1}
\begin{array}{l}
\displaystyle
\mathcal{L}_{\nu}(z)\,=\,\sum_{n=1}^\infty\,\frac{z^n}{n^{\nu}}\,,\qquad |z| < 1\,.
\end{array}
\end{equation}
 {For} $|z| \ge 1$, the~series in Equation~(\ref{Eq:KE.6.1}) can be analytically continuated to the whole complex plane, besides the
half-axis $\mathrm{Re}\{z\} > 1$.  

More importantly (for the scopes of the present paper),  functions $\mathcal{L}_{\nu}(z)$ are examples of  {Stieltjes function}. 
For the reader's convenience, the~main definitions of Stieltjes functions and of Stieltjes series will now be briefly summarized. 
Interested readers can found some more useful references for instance in Ref.~\cite{Caliceti/Meyer-Hermann/Ribeca/Surzhykov/Jentschura/2007}.
Moreover, the~ formalism of Allen~et~al.~\cite{Allen/Chui/Madych/Narcowich/Smith/Smith/1975} will be adopted.
Consider then an increasing real-valued  function $\mu(t)$ defined for $t\in[0,\infty]$, with~infinitely many points of increase. 
The measure $\mathrm{d}\mu$ is then positive on $[0,\infty]$. Assume all moments
\begin{equation}
\label{Eq:Stieltjes.0.1}
\displaystyle
\mu_m=\int_0^\infty\,t^m\,\mathrm{d}\mu\,,\qquad m \ge 0\,,
\end{equation}
to be finite. Then, the~formal power series
\begin{equation}
\label{Eq:Stieltjes.0.2}
\displaystyle
\sum_{m=0}^\infty\,\mu_m\,z^m\,,
\end{equation}
is called a { {Stieltjes series}}. Such series turns out to be asymptotic, for~ $z\to 0$,  
to the function $F(z)$  defined by 
\begin{equation}
\label{Eq:Stieltjes.0.3}
\displaystyle
F(z)\,=\,\int_0^\infty\,\frac{\mathrm{d}\mu}{1-zt}\,,
\end{equation}
%
which is called { {Stieltjes integral}}.
%
To give a significative example,  the~Stieltjes nature of  functions $\mathcal{L}_\nu$ will now be proved. 
To this end, it is sufficient to start from the following integral representation~\cite{NIST:DLMF} (25.12.11):
\begin{equation}
\label{Eq:Debye.1.1.2.2}
\begin{array}{l}
\displaystyle
\mathcal{L}_\nu(z)\,=\,
-\dfrac 1{\Gamma(\nu)}\,
\int_0^\infty\,
\dfrac{x^{\nu-1}\exp(-x)}{\exp(-x)-1/z\,}\,\d x,
\end{array}
\end{equation}
which, on~setting  $t=\exp(-x)$, gives  
\begin{equation}
\label{Eq:Debye.1.1.2.3}
\begin{array}{l}
\displaystyle
\mathcal{L}_\nu(z)\,=\,
z\,\int_0^1\,
\dfrac{\left(-\log t\right)^{\nu-1}/{\Gamma(\nu)}}{1-zt}\,
\mathrm{d}t\,,
\end{array}
\end{equation}
%
thus proving that the function $\mathcal{L}_\nu(z)/z$ is a Stieltjes integral exactly of the form~(\ref{Eq:Stieltjes.0.3}), 
with the measure
\begin{equation}
\label{Eq:Debye.1.1.2.4}
\mu(t)\,=\,
\left\{
\begin{array}{lr}
\displaystyle
\dfrac{\Gamma(\nu,-\log t)}{\Gamma(\nu)}\,, & \qquad \qquad 0\le t \le 1,\\
&\\
1\,& \qquad \qquad t>1\,,
\end{array}
\right.
\end{equation}
where symbols $\Gamma(\cdot)$ and $\Gamma(\cdot,\cdot)$ denote the gamma and the  { {incomplete}} gamma functions, respectively~\cite{NIST:DLMF}.

{The above described ``asymptotic connection'' between series~(\ref{Eq:KE.4}) and~(\ref{Eq:KE.6}) via \mbox{Equation~(\ref{Eq:KE.4.1})}} would 
at first sight suggest the function $\mathbb{S}(\epsilon;M)$ could  be Stieltjes too.
In the next section such a conjecture will definitely be~proved. 


%
\section{A Constructive Proof That \boldmath{$\mathbb{S}(\epsilon;M)$} Is a {Stieltjes} Series}
\label{Sec:StieltjesKapteyn}


We are now going to prove
the following theorem:
Function $\mathbb{S}(\epsilon;M)$ defined through the series~(\ref{Eq:KE.4}) is a Stieltjes function.


%
%
%
First of all, the~definition of $\mathbb{S}(\epsilon;M)$ given in Equation~(\ref{Eq:KE.4}) must be recast as follows:
\begin{equation}
\label{Eq:KE.4Bis}
\begin{array}{l}
\displaystyle
\mathbb{S}(\epsilon;M)\,=\,\sum_{n=1}^\infty\,
\frac{2\,J_n(n\,\epsilon)}n\exp(-\lambda n)\,z^n\,,
\end{array}
\end{equation}
with $z=\exp(\lambda+\mathrm{i}M)$. What we are going to show is that, similarly as for Equation~(\ref{Eq:Debye.1.1.2.3}),
$\mathbb{S}$ can be written as
\begin{equation}
\label{Eq:StieltjesKapteyn.1}
\begin{array}{l}
\displaystyle
\mathbb{S}(\epsilon;M)\,=\,
z\int_0^1\,\dfrac{\rho(t)\,\mathrm{d} t}{1-zt}\,,
\end{array}
\end{equation}
where the measure $\mathrm{d}\mu=\rho(t)\mathrm{d}t$ represents the proof goal. 
Accordingly, the~following infinite system must necessarily hold:
\begin{equation}
\label{Eq:StieltjesKapteyn.1.1}
\begin{array}{l}
\displaystyle
\frac{2\,J_n(n\,\epsilon)}n\exp(-\lambda n)\,=\
\int_0^1\,t^{n-1}\,\rho(t)\,\mathrm{d} t\,, \qquad\qquad n \ge 1\,.
\end{array}
\end{equation}
 {The} key tool of our proof is 
Watson's integral representation of $J_n(n\epsilon)$, namely~\cite{Watson/1917}
\begin{equation}
\label{Eq:WatsonJ}
\begin{array}{l}
\displaystyle
J_n(n\,\epsilon)\,=\,\dfrac 1\pi\,\int_0^\pi\,\exp[-n\,F(\theta;\epsilon)]\,\mathrm{d}\theta\,,\qquad\qquad 0\le\epsilon \le 1\,,
\end{array}
\end{equation}
where 
\begin{equation}
\label{Eq:WatsonJ.2}
\begin{array}{l}
\displaystyle
F(\theta;\epsilon)\,=\,
\log\dfrac{\theta+\sqrt{\theta^2-\epsilon^2\sin^2\theta}}{\epsilon\,\sin\theta}\,-\,\dfrac{\sqrt{\theta^2-\epsilon^2\sin^2\theta}}{\tan\theta}\,,\qquad
\theta\in[0,\pi]\,.
\end{array}
\end{equation}
 {Function} $F(\theta;\epsilon)$ has features which reveal very important for our scopes. 
One of them is 
\begin{equation}
\label{Eq:WatsonJ.2-1}
\begin{array}{l}
\displaystyle
F\left(0;\sqrt{1-\chi^2}\right)\,+\,\lambda\,=\,0\,,\qquad\qquad 0 \le \chi \le 1\,,
\end{array}
\end{equation}
which suggests Equation~(\ref{Eq:WatsonJ}) to be recast as
\begin{equation}
\label{Eq:WatsonJ.3}
\begin{array}{l}
\displaystyle
J_n(n\,\epsilon)\,\exp(-\lambda\,n)\,=\,\dfrac 1\pi\,\int_0^\pi\,\exp[-n\,G_\chi(\theta)]\,\mathrm{d}\theta\,,
\end{array}
\end{equation}
where function $G_\chi(\theta)$ is defined as $G_\chi(\theta)=F\left(\theta;\sqrt{1-\chi^2}\right)+\lambda$. 

\begin{figure}[!ht]
\includegraphics[width=6cm]{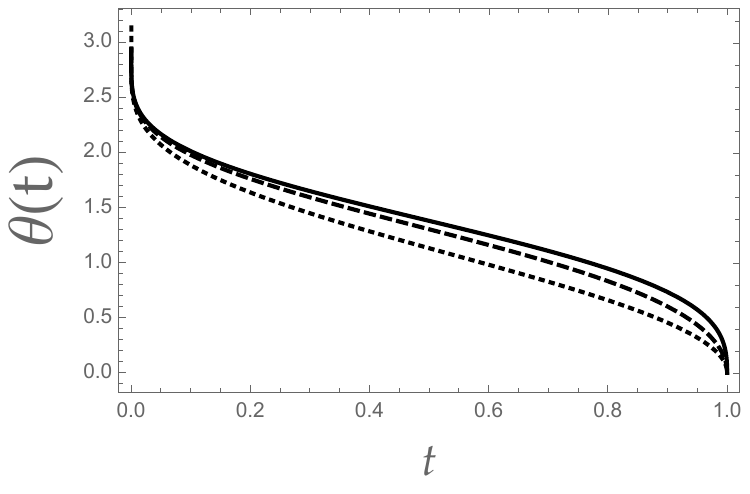}
\caption{Behaviour of the function $\theta(t)$, according to Equation~(\ref{Eq:WatsonJ.3.1}), for~$\chi=1/10$ (solid curve), $\chi=1/2$ (dashed curve), and~$\chi=1$ (dotted curve).}
\label{Fig:GraficiDiThetaOfTi}
\end{figure}

Another property of $F(\theta;\epsilon)$ is that 
$\partial F(\theta;\epsilon)/\partial\theta > 0$ for $\theta \in (0,\pi]$~\cite{Watson/1917}.
Then, in~order for Equation~(\ref{Eq:WatsonJ.3}) to be ``tuned'' with Equation~(\ref{Eq:StieltjesKapteyn.1.1}), 
the new integration variable \mbox{$t =\exp[-\,G_\chi(\theta)]$} is introduced, in~such a way that 
%
\begin{equation}
\label{Eq:WatsonJ.3.1}
\begin{array}{l}
\displaystyle
\dfrac{\mathrm{d} t}{\mathrm{d} \theta}\,=\,-\exp[-\,G_\chi(\theta)]\dfrac{\mathrm{d}G_\chi(\theta)}{\mathrm{d} \theta} \,<\, 0\,,
\end{array}
\end{equation}
for all values of $\theta$ except for $\theta=0$ and $\theta=\pi$, where $t'(\theta)=0$.
%
%
Accordingly,  function $t=t(\theta)$ can be uniquely inverted 
into the function $\theta=\theta(t)$, formally defined by 
\begin{equation}
\label{Eq:WatsonJ.3.1}
\begin{array}{l}
\displaystyle
\theta(t)=F^{-1}[-\log t\,-\,\lambda]\,,\qquad\qquad t \in[0,1]\,,
\end{array}
\end{equation}
where the symbol $F^{-1}$ denotes the inverse of $F(\theta;\sqrt{1-\chi^2})$ with respect to $\theta$. 
Although its explicit analytical expression cannot be found, $\theta(t)$ can be easily numerically computed
for any values of $t$ and $\chi$, for~instance on using Mathematica's native command {\tt InverseFunction}.
In Figure~\ref{Fig:GraficiDiThetaOfTi}, the~behaviour of $\theta(t)$, evaluated according to Equation~(\ref{Eq:WatsonJ.3.1}), is plotted  for different values of the aspect ratio $\chi$.

Now, we are going to prove that the function $\theta(t)$ actually coincides (apart some constant proportional factors), with~the density $\rho(t)$ into
Equation~(\ref{Eq:StieltjesKapteyn.1}). To~this end, it is sufficient to change the integration variable in Equation~(\ref{Eq:WatsonJ.3}) from $\theta$ to $t$, which  yields 
\begin{equation}
\label{Eq:WatsonJ.4}
\begin{array}{l}
\displaystyle
J_n(n\,\epsilon)\,\exp(-\lambda\,n)\,=\,-\dfrac 1\pi\,\int_0^1 t^n\,\theta'(t)\,\mathrm{d}t\,.
\end{array}
\end{equation}
 {On} performing the $t$-integration by parts we have
\begin{equation}
\label{Eq:WatsonJ.5}
\begin{array}{l}
\displaystyle
J_n(n\,\epsilon)\,\exp(-\lambda\,n)\,=\,-\dfrac 1\pi\,\left[t^n\,\theta(t)\right]_0^1\,+\,
\dfrac n\pi\,\int_0^1t^{n-1}\,\theta(t)\mathrm{d}t\,=\,
\dfrac n\pi\,\int_0^1t^{n-1}\,\theta(t)\mathrm{d}t\,,
\end{array}
\end{equation}
or, equivalently,
\begin{equation}
\label{Eq:WatsonJ.6}
\begin{array}{l}
\displaystyle
\dfrac{2J_n(n\,\epsilon)\,\exp(-\lambda\,n)}n\,=\,
\int_0^1\,t^{n-1}\,\dfrac{2\theta(t)}\pi\mathrm{d}t\,,\qquad\qquad n\ge 1\,,
\end{array}
\end{equation}
which, once compared to Equation~(\ref{Eq:StieltjesKapteyn.1.1}), leads to $\rho(t)=2\theta(t)/\pi$.

The uniqueness of the function $\rho(t)$ can further be proved by invoking Calerman's theorem for the Hausdorff problem,
i.e., by~showing that 
\begin{equation}
\label{Eq:Calerman}
\begin{array}{l}
\displaystyle
\sum_{m=0}^\infty\,\mu_m^{-1/{2m}}\,=\,+\infty\,.
\end{array}
\end{equation}
 {In} order to prove the divergence of the above series, it is worth starting from the following inequality, proved by Siegel~\cite{Siegel/1953}:
\begin{equation}
\label{Eq:Calerman.2}
\begin{array}{l}
\displaystyle
J_n(n\epsilon)\,\exp(-\lambda n) \,\le  1\,,\qquad\qquad n \ge 1\,,
\end{array}
\end{equation}
%
according to which we have
\begin{equation}
\label{Eq:Calerman.3}
\begin{array}{l}
\displaystyle
\dfrac n{J_n(n\epsilon)\,\exp(-\lambda n)} \,\ge  n\,,\qquad \qquad n \ge 1\,,
\end{array}
\end{equation}
and then
\begin{equation}
\label{Eq:Calerman.3}
\begin{array}{l}
\displaystyle
\lim_{n\to\infty}\left(\dfrac n{J_n(n\epsilon)\,\exp(-\lambda n)}\right)^{1/2n}\,\ge  \lim_{n\to\infty}n^{1/2n}\,=\,1\,,
\end{array}
\end{equation}
which leads to Equation~(\ref{Eq:Calerman}).
Our proof is now complete. 
$\mathbb{S}(\epsilon;M)$ {is} a Stieltjes function. 

\section{A New Integral Representation of KE's~Solution}
\label{Susec:NewIntegralRepresentation}

The proof given in Section~\ref{Sec:StieltjesKapteyn} is the main result of the present paper. 
Here, an~interesting byproduct of such proof is now illustrated. In~particular, 
the ``Stieltjesness'' of $\mathbb{S}(\epsilon;M)$ will be employed to conceive 
a new, up~to our knowledge, integral representation of the KE solution. 
To this end,  Equation~(\ref{Eq:StieltjesKapteyn.1}) is first recast as follows:
\begin{equation}
\label{Eq:SolvingKepler.1}
\begin{array}{l}
\displaystyle
\mathbb{S}(\epsilon;M)\,=\,
-\dfrac{2}\pi\,\int_0^1\,\dfrac{\theta(t)}{t- 1/z}\,\mathrm{d} t\,,
\end{array}
\end{equation}
so that formal partial integration gives
\begin{equation}
\label{Eq:SolvingKepler.2}
\begin{array}{l}
\displaystyle
\mathbb{S}(\epsilon;M)\,=\,
-\dfrac{2}\pi\,\left[\theta(t)\,\log\left(t- \dfrac 1z\right)\right]_0^1\,+\,
\dfrac{2}\pi\,\int_0^1\,\log\left(t- \dfrac 1z\right)\,\theta'(t)\,\mathrm{d} t\,.
\end{array}
\end{equation}
 {The} change of variable $t=\exp[-G_\chi(\theta)]$ then yields
\begin{equation}
\label{Eq:SolvingKepler.3}
\begin{array}{l}
\displaystyle
\mathbb{S}(\epsilon;M)\,=\,
2\,\log\left(-\dfrac 1z\right)\,\,-\,
\dfrac{2}\pi\,\int_0^\pi\,\log\left(\exp[-G_\chi(\theta)]- \dfrac 1z\right)\,\mathrm{d} \theta\,,
\end{array}
\end{equation}
which can be rearranged into a simpler form with a few work. First of all, we have
\begin{equation}
\label{Eq:SolvingKepler.3.1}
\begin{array}{l}
\displaystyle
\mathbb{S}(\epsilon;M)\,=\,
2\,\log\left(-\dfrac 1z\right)\,\,-\,
\dfrac{2}\pi\,\int_0^\pi\,\log\left(\dfrac{z\,\exp[-G_\chi(\theta)]-  1}z\right)\,\mathrm{d} \theta\,,
\end{array}
\end{equation}
or, equivalently,
\begin{equation}
\label{Eq:SolvingKepler.3.1.0}
\begin{array}{l}
\displaystyle
\mathbb{S}(\epsilon;M)\,=\,
2\,\log\left(-\dfrac 1z\right)\,\,-\,
\dfrac{2}\pi\,\int_0^\pi\,
\log\left(z\,\exp[-G_\chi(\theta)] - 1\right)\,\mathrm{d} \theta\,-\,
2\,\log \dfrac 1z\,{=}\,\\
\\
\displaystyle
\,{=}\,
2\mathrm{i}\pi
\,-\,
\dfrac{2}\pi\,\int_0^\pi\,\log\left(\exp[-F(\theta;\epsilon)+\mathrm{i}M]- 1\right)\,\mathrm{d} \theta\,.
\end{array}
\end{equation}
 {To} retrieve the KE solution, it is sufficient to recall that $S(\epsilon;M)=\mathrm{Im}\{\mathbb{S}(\epsilon;M)\}$, which gives at once
\vspace{-5pt}
\begin{equation}
\label{Eq:SolvingKepler.3.1.1.1}
\begin{array}{l}
\displaystyle
{S}(\epsilon;M)\,=\,
2\pi
\,+\,
\dfrac{\mathrm{i}}\pi\,
\int_0^\pi\,
\mathrm{d} \theta\,
\left\{
\log\left(\exp[-F(\theta;\epsilon)+\mathrm{i}M]- 1\right)\,\,-\,
\log\left(\exp[-F(\theta;\epsilon)-\mathrm{i}M]- 1\right)
\right\}\,,
\end{array}
\end{equation}
%
that can also be recast as
\begin{equation}
\label{Eq:SolvingKepler.3.1}
\begin{array}{l}
\displaystyle
{S}(\epsilon;M)\,=\,\dfrac{\mathrm{i}}\pi\,
\int_0^\pi\,\log\dfrac{1\,-\,\exp[-\,F(\theta;\epsilon)\,+\,\mathrm{i}M]}{1\,-\,\exp[-\,F(\theta;\epsilon)\,-\,\mathrm{i}M]}\,\mathrm{d} \theta\,,
\end{array}
\end{equation}
{{where the term} 
 $2\pi$ in Equation~(\ref{Eq:SolvingKepler.3.1.1.1}) must be removed, due to the phase wrapping of the principal branch of the natural logarithm, 
 $-\pi <  \arg\{\log z\} \le \pi$.}

Equation~(\ref{Eq:SolvingKepler.3.1}) is the other main result of the present paper. 
Compared with other, formally similar integral representations of the KE solution, for instance that in~\cite{Eisinberg/2010}, Equation~(\ref{Eq:SolvingKepler.3.1}) is expected to provide 
computational advantages due to the monotonic decreasing behaviour of the function $\exp[-F(\theta;\epsilon)]$ within the integration interval $\theta\in[0,\pi]$.

{
{Just to provide} an idea, a~single but significant  numerical experiment will now be carried out by using only native commands of { {Mathematica}}. 
A case of particular interest is that of nearly parabolic orbits around the periapsis, which correspond to $\epsilon \to 1$.
Accurate numerical solutions for such a critical regime can be found, for~instance, in~\cite{Tommasini/Olivieri/2022,Farnocchia/Cioci/Milani/2013}.
In particular, in~\cite{Farnocchia/Cioci/Milani/2013}, it was emphasized how solving the elliptic KE  for a given couple $(\epsilon,M)$ resulted in a bad conditioned 
problem in the neighbourhood of $(1,0)$. 
To~provide an example of the computational use of the new integral representation in Equation~(\ref{Eq:SolvingKepler.3.1}), we considered
the sole case of a parabolic orbit, i.e.,~ $\epsilon=1$.

{In Figure}~\ref{Fig:NumericalSimulation}, the relative error, evaluated with respect to the ``exact KE solution'', is plotted against $M\in(0,\pi)$. In~the above experiments, 
such an ``exact value''  was evaluated by solving Equation~(\ref{Eq:KE.1}) via Mathematica's native command {\tt FindRoot} with the parameter
{\tt WorkingPrecision} set to 50 and an initial guess of $\psi=\pi$.
Function $S(\epsilon;M)$ was evaluated by implementing Equation~(\ref{Eq:SolvingKepler.3.1.1.1}) using the native Mathematica command {\tt NIntegrate} with different degrees of accuracy, measured by the 
parameter {\tt WorkingPrecision}, which was set to 10 (a),  15  (b),  20 (c), and~ 25 (d).
\begin{figure}[!ht]
\begin{minipage}[t]{6cm}
{\includegraphics[width=6cm,angle=-0]{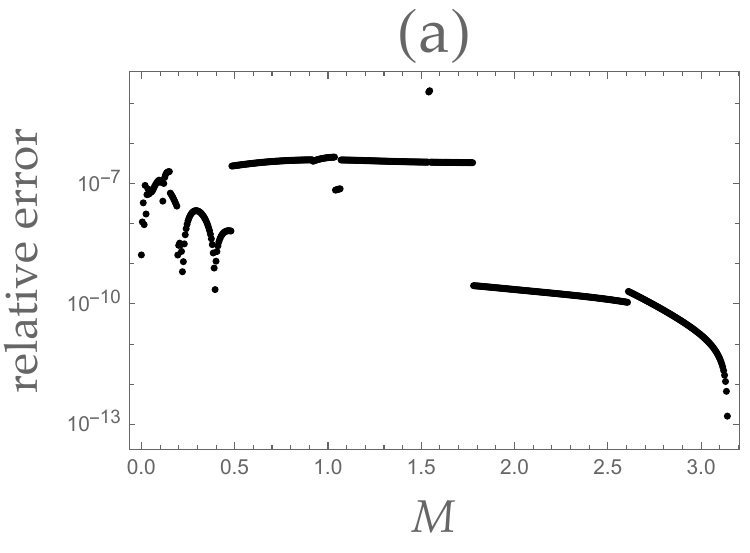}}
{\includegraphics[width=6cm,angle=-0]{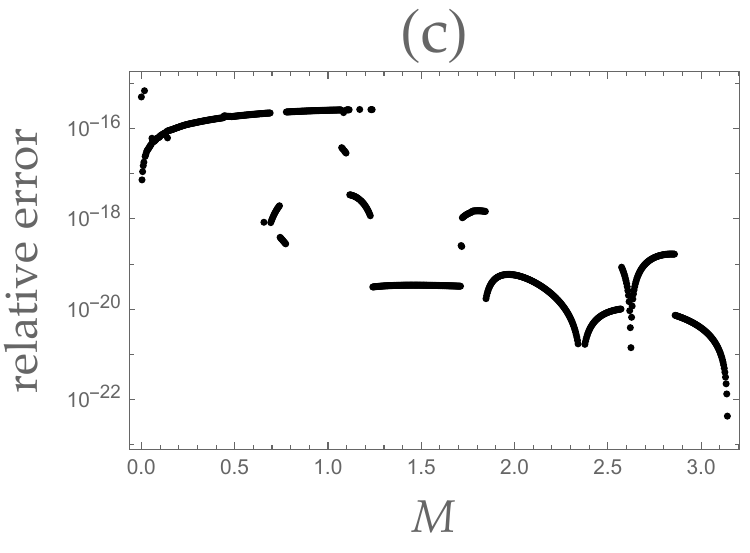}}
\end{minipage}
\begin{minipage}[t]{6cm}
{\includegraphics[width=6cm,angle=-0]{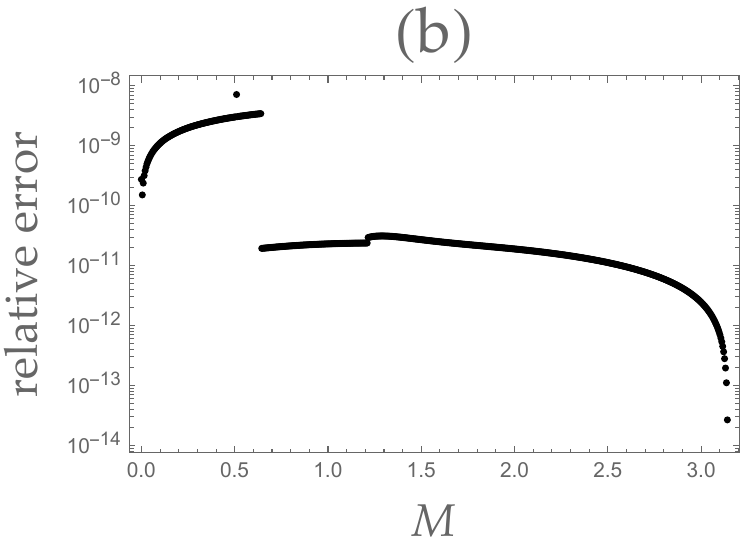}}
{\includegraphics[width=6cm,angle=-0]{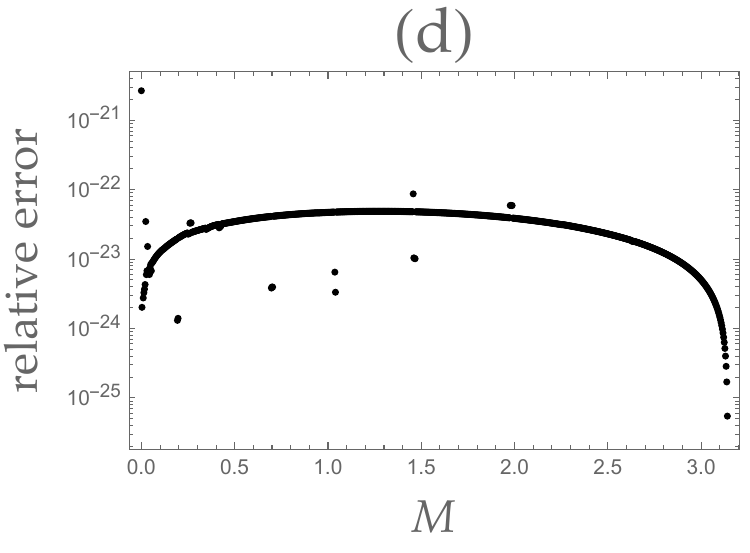}}
\end{minipage}
\caption{Behaviour of the relative error 
against $M\in(0,\pi)$. In~the above experiments, 
the ``exact value'' of the KE solution was evaluated by solving Equation~(\ref{Eq:KE.1}) via Mathematica's native command {\tt FindRoot} with the parameter
{\tt WorkingPrecision} set to 50 and the initial guess of \mbox{$\psi=\pi$}.
Function $S(\epsilon;M)$ was evaluated by implementing Equation~(\ref{Eq:SolvingKepler.3.1.1.1}) through the native Mathematica command {\tt NIntegrate} with different degrees of accuracy, measured by the 
parameter {\tt WorkingPrecision}, which was set to 10 (\textbf{a}),  15  (\textbf{b}),  20 (\textbf{c}), and~ 25 (\textbf{d}).
}
\label{Fig:NumericalSimulation}
\end{figure}
}

\section{Discussions}
\label{Sec:Discussions}

Here, the classical Bessel solution of Kepler's equation was revisited 
from a new perspective, offered by the beautiful theory of the Stieltjes series. 
In particular, after~introducing a ``complexified''  version of the original Kapteyn series~(\ref{Eq:KE.3}), its Stieltjes nature was mathematically proven. 
Our principal tool was Watson's integral representation of Bessel functions.
As a potentially interesting by-product of our analysis,~to our knowledge, a new
integral representation of the KE solution  was~achieved.

Exploring the Stieltjes nature of Equation~(\ref{Eq:KE.4}) could also reveal new interesting aspects concerning the more general framework of the Kapteyn series.
The need to develop new methods to achieve the summability of the Kapteyn series of both the first and second kind
have already been invoked in the recent past~\cite{Tautz/Lerche/Dominici/2011}. 
In this respect, we believe the results presented here could be of some~help.

Just to gain an idea, consider again the Kapteyn series $\displaystyle\sum_{m\ge 1}\,\dfrac{z^m}m\,J_m(m\epsilon)$,
but suppose that  $|z| > \exp(-\lambda)$. In~that case, the series diverges.
However, the~analysis carried out in Section~\ref{Sec:StieltjesKapteyn} is also still valid for $z$ belonging 
to the whole complex plane, provided that $|\mathrm{arg}(z)|< \pi$.
In other words, our  Kapteyn series can be continued to a Stieltjes function according to
\begin{equation}
\label{Eq:StieltjesKapteynSeries}
\begin{array}{l}
\displaystyle
\displaystyle\sum_{m\ge 1}\,\dfrac{z^m}m\,J_m(m\epsilon)\,\sim\,
\mathrm{i}\pi
\,-\,
\dfrac{1}\pi\,\int_0^\pi\,\log\left(z\,\exp[-F(\theta;\epsilon)]- 1\right)\,\mathrm{d} \theta\,,\quad |\mathrm{arg}(z)|< \pi\,.
\end{array}
\end{equation}
which lives, for~$\epsilon \in [0,1)$, on~the whole complex plane, besides the infinite interval $(1,\infty)$. 

Proving that a divergent power series is a Stieltjes series also implies that such a series would be Pad\'{e} summable.
To quote an important example, in~\cite{Bender/Weniger/2001}, numerical evidence for the factorially divergent perturbation expansion for  
energy eigenvalues of the PT-symmetric Hamiltonian $H(\lambda) =p^2+1/4x^2+\mathrm{i} \lambda x^3$ being a Stieltjes series and thus Pad\'e 
summable, were provided. 
 Such conjecture was later rigorously proven in~\cite{Grecchi/Maioli/Martinez/2009}.
The Pad\'e summability of our  Kapteyn series also implies that suitable resummation algorithms, 
the most well-known being Wynn's $\epsilon$ algorithm, can successfully be employed to retrieve the correct value provided in Equation~(\ref{Eq:StieltjesKapteynSeries}).
Pad\'e approximants remain the most favourite mathematical tool to resume divergent and slowly convergent series in theoretical physics. 
 This is because a solid
convergence theory has been established for them, especially in the case of  Stieltjes asymptotic series~\cite{Bender/Orszag/1978,Baker/Graves-Morris/1996,Widder/1938}. 
In particular, if~the input data are the partial sums of a Stieltjes series, it can be rigorously proven that certain subsequences of the Pad\'{e} 
table converge to the value of the corresponding Stieltjes function~\cite{Baker/Graves-Morris/1996,Brezinski/1996} (Chapter 5). 
A typical feature of Wynn's $\epsilon$ algorithm is that only the input of the numerical values of a finite substring of the partial sequence of the series is required to achieve a meaningful 
resummation. 

In the case of the Stieltjes series, however, important additional structural information about the behaviour of the series terms is available. For~example, it is known that the magnitudes of the 
truncation errors are bounded by the first term neglected in the partial sum, and~they also have the same sign patterns. 
In the seminal 1973 paper by Levin~\cite{Levin/1973}, a new class of sequence transformations was introduced that used such valuable { {a priori} 
} structural information 
to improve the efficiency and convergence speed of the transformed sequences. 
For a general discussion on the construction principles of Levin-type sequence transformations, readers are encouraged to go through the paper by 
Weniger~\cite{Weniger/1989},  as well as through Ref.~\cite{Borghi/Weniger/2015}, which contains a slightly upgraded and formally improved version of the former. 
To provide a simple example, Table~\ref{Tab.Finale.KS} shows the performance of a particular Levin-type transformation, the~so-called Weniger $\delta$-transformation~\cite{Weniger/1989}, 
in the resummation of the Kapteyn series on the left side of Equation~(\ref{Eq:StieltjesKapteynSeries}) for $\epsilon=9/10$ and $z=10\,\exp(\mathrm{i}\pi/3)$. 

The convergence to  the value obtained by numerically computing  the integral on the right side of Equation~(\ref{Eq:StieltjesKapteynSeries}) is evident.
\begin{table}[!ht]
\centerline{
    \begin{tabular}{|c|c|c|}
    \hline
  order & Partial sum sequence			 & Weniger $\delta$-transformation 			\\ \hline \hline
 1 & 2.02+3.51 i                         			&  0.112240 + 1.211289 i 	\\
 10 & (4.4 - 10. i) $\times 10^8$  		& -1.003096 +1.238166 i 	\\
 20 & (-3.1 + 32. i)  $\times 10^{18}$ 		& -1.001839 + 1.238763 i	\\
 30 & (7.7 +10. i) $\times 10^{27}$ 		& -1.001838 + 1.238765 i 	\\
 \ldots & \ldots  						& \ldots  				\\ \hline
  \hline
    \end{tabular}
}
\caption{ {Resummation,} 
 via Weniger $\delta$-transformation~\cite{Weniger/1989}, of~the (divergent) Kapteyn series $\displaystyle\sum_{m\ge 1}\,\dfrac{z^m}m\,J_m(m\epsilon)$, 
evaluated  for $z=10\,\exp(\mathrm{i}\pi/3)$. 
The value obtained by numerically computing the integral into the right side of Equation~(\ref{Eq:StieltjesKapteynSeries}) is: $-$1.001838\ldots + 1.238765\ldots i.}
\label{Tab.Finale.KS}
\end{table}
{{Finally,} 
 before~concluding, it is worth illustrating at least one experiment of the use of the Weniger transformation to solve Kepler's equation.
Once again, the~parabolic case will be analysed, but~now only ``small values'' of $M$, namely less than $1/10$, will be considered. In~Figure~\ref{Fig:NumericalSimulationWeniger},
the relative error is still plotted  {versus} 
 $M\in \left[\dfrac 1{1000},\dfrac 1{10}\right]$. Again, the~``exact solution'' was evaluated by solving Equation~(\ref{Eq:KE.1}) via Mathematica's  
command {\tt FindRoot} with the parameter {\tt WorkingPrecision} set to 50 and an initial guess of $\psi=\pi$, but~now the function $S(\epsilon;M)$ is thought of as the imaginary part of 
$\mathbb{S}(\epsilon;M)$. The~latter is computed, as~done in Table~\ref{Tab.Finale.KS}, via the Weniger $\delta$-transformation with different orders. 
\begin{figure}[!ht]
{\includegraphics[width=8cm,angle=-0]{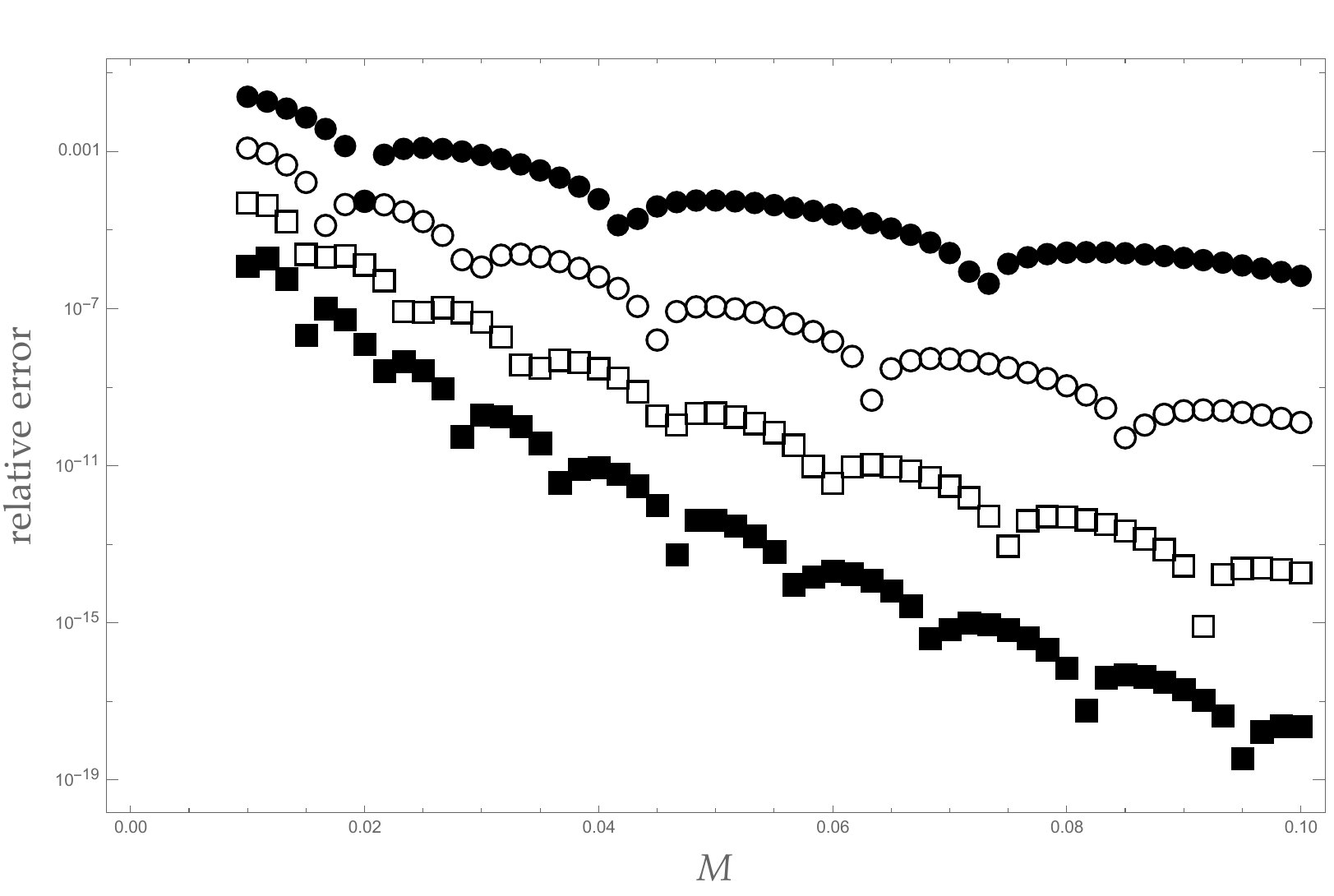}}
\caption{The same as in Figure~\ref{Fig:NumericalSimulation}, but~for $M\in \left[\dfrac 1{1000},\dfrac 1{10}\right]$. Note that now the function $S(\epsilon;M)$ is thought of as the imaginary part of 
$\mathbb{S}(\epsilon;M)$, which is computed, similarly to that in Table~\ref{Tab.Finale.KS}, via the Weniger $\delta$-transformation
with an order of 20 (black circles), 30 (open circles), 40 (open squares), and~50 (black squares).}
\label{Fig:NumericalSimulationWeniger} 
\end{figure}

 }

\newpage
\section{Conclusions}

\begin{quotation}
{\em \noindent  {In common with} 
 almost any scientific problem which achieves a certain longevity and whose literature exceeds a certain critical mass, the~Kepler problem 
has acquired luster and allure for the modern practitioner. Any new technique for the treatment of trascendental equations should be applied to this illustrious case; any new 
insight, however slight, lets its conceiver join an eminent list of contributors.
}
\end{quotation}

Nothing can illustrate the fascination that Kepler's equation has held for nearly 400 years better than the above paragraph, which appears in the introduction to Colwell's book.
Kepler's problem continues to inspire mathematicians, scientists, and~engineers to search for further contributions to the subject. 
In the present paper, the~Bessel solution of KE has been studied from the perspective of Stieltjes function theory. 
In particular, it has been proven that the Kapteyn series~(\ref{Eq:KE.3}) obtained by the ``complexification'' of the Bessel series solution~(\ref{Eq:KE.2}) is a Stieltjes series.
Moreover, thanks to our constructive proof, to~our knowledge, a new integral representation of the KE solution has been obtained. 
More importantly, the~limit of validity of such a representation extends far beyond that fixed by the convergence circle of the Kapteyn series~itself. 

Stieltjes series are strictly connected with Pad\'e summability via a well established convergence theory. Divergent Stieltjes series can be resummed to 
the corresponding Stieltjes integral values using Pad\'e approximants.
In the last 40 years, Levin-type transformations have been shown to possess retrieving capabilities considerably superior to those of Pad\'e, as~far as the resummation of the Stieltjes series
is concerned. At~present, the~main theoretical drawback of Levin-type transformations is the lack of a rigorous convergence theory, like that available for Pad\'e.
Thus, we also hope that what is contained in the present paper could provide a further contribution to the construction of such a convergence theory, also in order to provide Levin-type transformations the scientific recognition they certainly deserve, but~that, unfortunately, in~our opinion, still seems too far~away.

\section*{Acknowledgements}
I wish to thank Gabriella Cincotti and Turi Maria Spinozzi for their useful comments and help.

\noindent
Dedicated to Maurizio Giura, on his eighty-fifth birthday.


\begin{thebibliography}{999}


\bibitem[Colwell(1993)]{Colwell/1993}
Colwell, P.
\newblock {\em Solving Kepler's Equation Over Three Centuries}; Willmann-Bell:
  Richmond,  {VA, USA,} 
 1993.

\bibitem[Ibrahim and Saleh(2019)]{Ibrahim/Saleh/2019}
Ibrahim, R.; Saleh, A.R.
\newblock Re-evaluation solution methods for Kepler's equation of an elliptical
  orbit.
\newblock {\em Iraqi J. Sci.} {\bf 2019}, {\em 60},~2269--2279.

\bibitem[Calvo et~al.(2019)Calvo, Elipe, Montijano, and
  R\'andez]{Calvo/Elipe/Montijano/Randez/2019}
Calvo, M.; Elipe, A.; Montijano, J.; R\'andez, L.
\newblock A monotonic starter for solving the hyperbolic Kepler equation by
  Newton method.
\newblock {\em Celest. Mech. Dyn. Astron.} {\bf 2019}, {\em
  131},~18.

\bibitem[Tommasini and Olivieri(2020)]{Tommasini/Olivieri//2020}
Tommasini, D.; Olivieri, D.
\newblock Fast switch and spline scheme for accurate inversion of nonlinear
  functions: The new first choice solution to Kepler's equation.
\newblock {\em Appl. Math. Comput.} {\bf 2020}, {\em
  364},~124677.

\bibitem[Abubekerov and Gostev(2020)]{Abubekerov/Gostev/2020}
Abubekerov, M.K.; Gostev, N.Y.
\newblock Solution of Kepler's Equation with Machine Precision.
\newblock {\em Astron. Rep.} {\bf 2020}, {\em 64},~1060--1066.

\bibitem[Sacchetti(2020)]{Sacchetti/2020}
Sacchetti, A.
\newblock Francesco Carlini: Kepler's equation and the asymptotic solution to
  singular differential equations.
\newblock {\em Hist. Math.} {\bf 2020}, {\em 53},~1--32.

\bibitem[Tommasini and Olivieri(2020)]{Tommasini/Olivier/2020}
Tommasini, D.; Olivieri, D.
\newblock Fast Switch and Spline Function Inversion Algorithm with Multistep
  Optimization and k-Vector Search for Solving Kepler's Equation in Celestial
  Mechanics.
\newblock {\em Mathematics} {\bf 2020}, {\em 8},~2017.

\bibitem[Zechmeister(2021)]{Zechmeister/2021}
Zechmeister, M.
\newblock Solving Kepler'ss equation with CORDIC double iterations.
\newblock {\em Mon. Not. R. Astron. Soc.} {\bf 2021}, {\em 500},~109--117.

\bibitem[Gonz\'alez-Gaxiola and
  Hern\'andez-Linares(2021)]{Gonzalez/Hernandez/2021}
Gonz\'alez-Gaxiola, O.; Hern\'andez-Linares, S.
\newblock An Efficient Iterative Method for Solving the Elliptical Kepler's
  Equation.
\newblock {\em Int. J. Appl. Comput. Math.} {\bf 2021}, {\em 7},~42.


\bibitem[Tommasini(2021)]{Tommasini/2021}
Tommasini, D.
\newblock Bivariate Infinite Series Solution of Kepler's Equation.
\newblock {\em Mathematics} {\bf 2021}, {\em 9},~785.

\bibitem[Philcox et~al.(2021)Philcox, Goodman, and
  Slepian]{Philcox/Goodman/Slepian/2021}
Philcox, O.H.E.; Goodman, J.; Slepian, Z.
\newblock Kepler's Goat Herd: An exact solution to Kepler's equation for
  elliptical orbits.
\newblock {\em Mon. Not. R. Astron. Soc.} {\bf 2021},
  {\em 506},~6111--6116.

\bibitem[Tommasini(2022)]{Tommasini/Olivieri/2022}
Tommasini, D.; Olivieri, D.
\newblock Two fast and accurate routines for solving the elliptic Kepler equation for all values of the eccentricity and mean anomaly.
\newblock {\em Astron. Astrophys.} {\bf 2022}, {\em 658},~A196.

\bibitem[Zhang et~al.(2022)Zhang, Bian, and Li]{Zhang/Bian/Li/2022}
Zhang, R.; Bian, S.; Li, H.
\newblock Symbolic iteration method based on computer algebra analysis for
  Kepler's equation.
\newblock {\em Sci. Rep.} {\bf 2022}, {\em 12},~2957.

\bibitem[Wu et~al.(2023)Wu, Zhou, Lim, and Zhong]{Zhou/Lim/Zhong/2023}
Wu, B.; Zhou, Y.; Lim, C.W.; Zhong, H.
\newblock A new solution approach via analytical approximation of the elliptic
  Kepler equation.
\newblock {\em Acta Astronaut.} {\bf 2023}, {\em 202},~303--310.

\bibitem[Vavrukh et~al.(2023)Vavrukh, Dzikovskyi, and
  Stelmakh]{Vavrukh/Dzikovskyi/Stelmakh/2023}
Vavrukh, M.; Dzikovskyi, D.; Stelmakh, O.
\newblock Analytical images of Kepler's equation solutions and their
  applications.
\newblock {\em Math. Model. Comput.} {\bf 2023}, {\em 10},~351--358.

\bibitem[Calvo et~al.(2023)Calvo, Elipe, and R\'andez]{Calvo/Elipe/Randez/2023}
Calvo, M.; Elipe, A.; R\'andez, L.
\newblock On the integral solution of elliptic Kepler's equation.
\newblock {\em Celest. Mech. Dyn. Astron.} {\bf 2023}, {\em
  135},~26.

\bibitem[Brown(2023)]{Brown/2023}
Brown, M.T.
\newblock An improved cubic approximation for K epler's equation.
\newblock {\em Mon. Not. R. Astron. Soc.} {\bf 2023},
  {\em 525},~57--66.

\bibitem[Kapteyn(1893)]{Kapteyn/1893}
Kapteyn, W.
\newblock Researches sur les functions de Fourier-Bessel.
\newblock {\em Ann. Sci. L'Ecole Norm. Sup. Ser.} {\bf 1893}, {\em
  3},~91--122.

\bibitem[Dominici(2007)]{Dominici/2007}
Dominici, D.
\newblock A new Kapteyn series.
\newblock {\em Integral Transform. Spec. Funct.} {\bf 2007}, {\em
  18},~409--418.

\bibitem[Lerche and Tautz(2007)]{Lerche/Tautz/2007}
Lerche, I.; Tautz, R.
\newblock A note on summation of kapteyn series in astrophysical problems.
\newblock {\em Astrophys. J.} {\bf 2007}, {\em 665},~1288--1291.

\bibitem[Pog\'any(2007)]{Pogany/2007}
Pog\'any, T.K.
\newblock Convergence of generalized Kapteyn expansion.
\newblock {\em Appl. Math. Comput.} {\bf 2007}, {\em 190},~1844--1847.

\bibitem[Lerche and Tautz(2008)]{Lerche/Tautz/2008}
Lerche, I.; Tautz, R.
\newblock Kapteyn series arising in radiation problems.
\newblock {\em J. Phys. A Math. Theor.} {\bf 2008},
  {\em 41},  {035202.} 


\bibitem[Lerche et~al.(2009)Lerche, Tautz, and
  Citrin]{Lerche/Tautz/Citrin/2009}
Lerche, I.; Tautz, R.; Citrin, D.
\newblock Terahertz-sideband spectra involving Kapteyn series.
\newblock {\em J. Phys. A Math. Theor.} {\bf 2009},
  {\em 42},  {365206}.

\bibitem[Eisinberg et~al.(2010)Eisinberg, Fedele, Ferrise, and
  Frascino]{Eisinberg/2010}
Eisinberg, A.; Fedele, G.; Ferrise, A.; Frascino, D.
\newblock On an integral representation of a class of Kapteyn (Fourier-Bessel)
  series: Kepler's equation, radiation problems and Meissel's expansion.
\newblock {\em Appl. Math. Lett.} {\bf 2010}, {\em 23},~1331--1335.

\bibitem[Tautz et~al.(2011)Tautz, Lerche, and
  Dominici]{Tautz/Lerche/Dominici/2011}
Tautz, R.C.; Lerche, I.; Dominici, D.
\newblock Methods for summing general Kapteyn series.
\newblock {\em J. Phys. A Math. Theor.} {\bf 2011}, {\em
  44}, 385202.

\bibitem[Baricz et~al.(2011)Baricz, Jankov, and
  Pog\'any]{Baricz/Jankov/Pogany/2011}
Baricz, A.; Jankov, D.; Pog\'any, T.K.
\newblock Integral representation of first kind Kapteyn series.
\newblock {\em J. Math. Phys.} {\bf 2011}, {\em 52},~043518.

\bibitem[Nikishov(2014)]{Nikishov/2014}
Nikishov, A.I.
\newblock Kapteyn series and photon emission.
\newblock {\em Bull. Lebedev Phys. Inst.} {\bf 2014}, {\em
  41},~332--338.

\bibitem[Baricz et~al.(2017)Baricz, Masirevic, and
  Pog\'any]{Baricz/Dragana/Masirevic/Pogany/2017}
Baricz, A.; Masirevic, D.J.; Pog\'any, T.K.
\newblock Kapteyn Series.
\newblock {\em Lect. Notes Math.} {\bf 2017}, {\em 2207},~87--111.

\bibitem[Xue et~al.(2019)Xue, Li, Man, Xing, Liu, Li, and
  Wu]{Xue/Li/Man/Xing/Liu/Li/Wu/2019}
Xue, X.; Li, Z.; Man, Y.; Xing, S.; Liu, Y.; Li, B.; Wu, Q.
\newblock Improved Massive MIMO RZF Precoding Algorithm Based on Truncated
  Kapteyn Series Expansion.
\newblock {\em Information} {\bf 2019}, {\em 10},~136.

\bibitem[Bornemann(2023)]{Bornemann/2023}
Bornemann, F.
\newblock A Jentzsch-Theorem for Kapteyn, Neumann and General Dirichlet Series.
\newblock {\em Comput. Methods Funct. Theory} {\bf 2023}, {\em
  23},~723--739.

\bibitem[Bender and Orszag(1978)]{Bender/Orszag/1978}
Bender, C.M.; Orszag, S.A.
\newblock {\em Advanced Mathematical Methods for Scientists and Engineers};
  McGraw-Hill: New York,   {NY, USA,} 
 1978.

\bibitem[Baker and Graves-Morris(1996)]{Baker/Graves-Morris/1996}
Baker, Jr., G.A.; Graves-Morris, P.
\newblock {\em {P}ad\'{e} Approximants}, 2nd ed.; Cambridge U. P.: Cambridge,  {UK,}
  1996.

\bibitem[Widder(1938)]{Widder/1938}
Widder, D.V.
\newblock The {S}tieltjes transform.
\newblock {\em Trans. Am. Math. Soc.} {\bf 1938}, {\em 43},~7--60.

\bibitem[Brezinski(1996)]{Brezinski/1996}
Brezinski, C.
\newblock Extrapolation algorithms and {P}ad\'{e} approximations: A historical
  survey.
\newblock {\em Appl. Numer. Math.} {\bf 1996}, {\em 20},~299--318.

\bibitem[Orlando et~al.(2018)Orlando, Farina, Zarro, and
  Terra]{Orlando/Farina/Zarro/Terra/2018}
Orlando, F.; Farina, C.; Zarro, C.; Terra, P.
\newblock Kepler's equation and some of its pearls.
\newblock {\em Am. J. Phys.} {\bf 2018}, {\em 86},~11.

\bibitem[Watson(1944)]{Watson/1944}
Watson, G.N.
\newblock {\em A {T}reatise on the {T}heory of {B}essel {F}unctions}; Cambridge
  University Press: Cambridge, UK,  1944.

\bibitem[Caliceti et~al.(2007)Caliceti, Meyer-Hermann, Ribeca, Surzhykov, and
  Jentschura]{Caliceti/Meyer-Hermann/Ribeca/Surzhykov/Jentschura/2007}
Caliceti, E.; Meyer-Hermann, M.; Ribeca, P.; Surzhykov, A.; Jentschura, U.D.
\newblock From useful algorithms for slowly convergent series to physical
  predictions based on divergent perturbative expansions.
\newblock {\em Phys. Rep.} {\bf 2007}, {\em 446},~1--96.

\bibitem[Allen et~al.(1975)Allen, Chui, Madych, Narcowich, and
  Smith]{Allen/Chui/Madych/Narcowich/Smith/Smith/1975}
Allen, G.D.; Chui, C.K.; Madych, W.R.; Narcowich, F.J.; Smith, P.W.
\newblock Pad{\'e} approximation of Stieltjes series.
\newblock {\em J. Approx. Theory} {\bf 1975}, {\em 14},~302--316.

\bibitem[{\relax DLMF}()]{NIST:DLMF}
Olver, F.W.J.;  {Olde Daalhuis}, A.B.; Lozier, D.W.; Schneider, B.I.; 
  Boisvert, R.F.; Clark, C.W.; Miller, B.R.; Saunders, B.V.; Cohl, H.S.; 
  McClain, M.A. (eds.) {\it NIST Digital Library of Mathematical Functions}. Release 1.1.3 of 2021-09-15.  2022.
\newblock Available online: \url{http://dlmf.nist.gov/}  

\bibitem[Watson(1917)]{Watson/1917}
Watson, G.N.
\newblock Bessel functions and Kapteyn series.
\newblock {\em Proc. Lond. Math. Soc.} {\bf 1917}, {\em 16},~150--174.

\bibitem[Siegel(1953)]{Siegel/1953}
Siegel, K.M.
\newblock An inequality involving Bessel functions of argument nearly equal to
  their order.
\newblock {\em Proc. Am. Math. Soc.} {\bf 1953},
  {\em 4},~858--859.

\bibitem[Farnocchia(2013)]{Farnocchia/Cioci/Milani/2013}
Farnocchia, D.; Cioci, D.B.; Milani, A. 
\newblock Robust resolution of Kepler's equation in all eccentricity regimes.
\newblock {\em Celest. Mech. Dyn. Astron.} {\bf 2013}, {\em 116}, ~21--34.

\bibitem[Bender and Weniger(2001)]{Bender/Weniger/2001}
Bender, C.M.; Weniger, E.J.
\newblock Numerical evidence that the perturbation expansion for a
  non-{H}ermitian $\mathcal{PT}$-symmetric {H}amiltonian is {S}tieltjes.
\newblock {\em J. Math. Phys.} {\bf 2001}, {\em 42},~2167--2183.

\bibitem[Grecchi et~al.(2009)Grecchi, Maioli, and
  Martinez]{Grecchi/Maioli/Martinez/2009}
Grecchi, V.; Maioli, M.; Martinez, A.
\newblock Pad\'{e} summability of the cubic oscillator.
\newblock {\em J. Phys. A} {\bf 2009}, {\em 42},~{425208}.

\bibitem[Levin(1973)]{Levin/1973}
Levin, D.
\newblock Development of non-linear transformations for improving convergence
  of sequences.
\newblock {\em Int. J. Comput. Math. B} {\bf 1973}, {\em 3},~371--388.

\bibitem[Weniger(1989)]{Weniger/1989}
Weniger, E.J.
\newblock Nonlinear sequence transformations for the acceleration of
  convergence and the summation of divergent series.
\newblock {\em Comput. Phys. Rep.} {\bf 1989}, {\em 10},~189--371.

\bibitem[Borghi and Weniger(2015)]{Borghi/Weniger/2015}
Borghi, R.; Weniger, E.J.
\newblock Convergence analysis of the summation of the factorially divergent
  {E}uler series by {P}ad\'{e} approximants and the delta transformation.
\newblock {\em Appl. Numer. Math.} {\bf 2015}, {\em 94},~149--178.

\end{thebibliography}
\end{document}